\documentclass[aps,twocolumn,prb,eqsecnum,showpacs]{revtex4}
\usepackage{graphicx}
\begin{document}
\draft
\preprint{\today}
\title{Fermionic versus bosonic descriptions of one-dimensional
       spin-gapped antiferromagnets}
\author{Shoji Yamamoto and Kei-ichi Funase}
\address{Division of Physics, Hokkaido University,
         Sapporo 060-0810, Japan}
\date{\today}
\begin{abstract}
In terms of spinless fermions and spin waves, we describe magnetic
properties of a spin-$\frac{1}{2}$ ferromagnetic-antiferromagnetic
bond-alternating chain which behaves as a Haldane-gap antiferromagnet.
On one hand, we employ the Jordan-Wigner transformation and treat the
fermionic Hamiltonian within the Hartree-Fock approximation.
On the other hand, we employ the Holstein-Primakoff transformation and
modify the conventional spin-wave theory so as to restore the sublattice
symmetry.
We calculate the excitation gap, the specific heat, the magnetic
susceptibility, magnetization curves, and the nuclear spin-lattice
relaxation rate with varying bond alternation.
These schemes are further applied to a bond-alternating tetramerized chain
which behaves as a ferrimagnet.
The fermionic language is particularly stressed as a useful tool to
investigate one-dimensional spin-gapped antiferromagnets, while the
bosonic one works better for ferrimagnets.
\end{abstract}
\maketitle

\section{Introduction}

   Haldane \cite{H464,H1153} sparked renewed interest in one-dimensional
Heisenberg antiferromagnets, predicting that their low-energy structures
should qualitatively vary according as the constituent spins are integral
or fractional.
A magnetic excitation gap immediately above the ground state, which is
referred to as the Haldane gap, was indeed observed in
quasi-one-dimensional spin-$1$ Heisenberg antiferromagnets such as
CsNiCl$_3$ \cite{B371} and Ni(C$_2$H$_8$N$_2$)$_2$NO$_2$(ClO$_4$).
\cite{R945,K86}
A rigorous example of such a massive phase was also given theoretically.
\cite{A799,A477}
Significant numerical efforts \cite{Y3348,S493,Y102,W14529,T047203} were
dovoted to detecting the Haldane gap in the higher-spin systems.
Competition between massive and massless phases in low-dimensional
quantum magnets was extensively studied especially by the
nonlinear-sigma-model quantum field theory
\cite{A397,A409,S3299,S3443,F14709,F2530,F8799,F398,K622,T5124,T8863}
and a wide variety of spin gaps$-$energy gaps in magnetic excitation
spectra$-$were further predicted.
There followed stimulative findings, including quantized plateaux in
zero-temperature magnetization curves, \cite{O1984,T103,C5126} gap
formation in coupled spin chains \cite{D5744,G8901} and the dramatic
crossover from one- to two-dimensional quantum antiferromagnets,
\cite{D618} and an antiferromagnetic excitation gap with a ferromagnetic
background. \cite{K3336,T5355,M68,T15189}

   Besides the sigma-model study, analytic approaches played a crucial
role in revealing the nature of Haldane-gap antiferromagnets.
The valence-bond-solid model \cite{A397,A409} stimulated considerable
interest in matrix-product representation
\cite{F381,K281,K293,T6443,T1639,Y157,Y1795}
of the Haldane phase.
The Lieb-Schultz-Mattis theorem \cite{L407} was generalized \cite{O1984}
to clarify a mechanism for gap formation in a magnetic field.
However, these arguments were essentially restricted to the ground-state
behavior and can hardly be extended to finite-temperature properties.
Numerical tools such as quantum Monte Carlo and density-matrix
renormalization-group techniques are indeed useful for such a purpose, but
an analytic strategy is still indispensable to low-temperature
thermodynamics especially of spin-gapped antiferromagnets, where grand
canonical sampling is hardly feasible numerically.
Then we are led to describe massive spin chains in terms of conventional
languages such as the Jordan-Wigner fermions and the Holstein-Primakoff
spin waves.

   The Jordan-Wigner transformation is an efficient approach to
low-dimensional quantum magnetism.
Spin-$\frac{1}{2}$ arrays with uniform \cite{B685} and alternating
\cite{B684,K687,Z181} antiferromagnetic exchange interactions between
nearest neighbors were thus investigated and their energy structures,
magnetization curves, and thermodynamic properties were
indeed revealed well.
Two-leg antiferromagnetic spin ladders were also discussed within this
scheme \cite{A6136,A6233} and the interchain-coupling effect on the
lowest-lying excitation was elucidated.
More refined fermionization was further proposed for coupled spin chains.
Ordering spinless fermions along a snake-like path, Dai and Su
\cite{D964} succeeded in interpreting massive and massless excitations
with varying number of the ladder legs.
Their idea was generalized to investigate zero-temperature magnetization
curves \cite{H1607} and thermodynamic quantities. \cite{H549}
In such circumstances, we consider fermionizing an spin-$\frac{1}{2}$
ferromagnetic-antiferromagnetic bond-alternating chain, which converges
to the spin-$1$ antiferromagnetic Heisenberg chain as the ferromagnetic
coupling tends to infinity and therefore reproduces many of observations
common to Haldane-gap antiferromagnets.

   Bosonic theory has significantly been developed for one-dimensional
quantum magnets in recent years.
While the Schwinger-boson mean-field theory is unable to distinguish
fractional-spin chains from integral-spin ones, it is still useful in
predicting the asymptotic dependence of the Haldane gap on spin quantum
number \cite{A316} and explaining quantum phase transitions of Haldane-gap
antiferromagnets in a field. \cite{X054419}
The Schwinger-boson representation was further applied to ferrimagnetic
spin chains \cite{W1057,Y064426} and ladders. \cite{C915}
It was a major breakthrough leading to the subsequent development of the
spin-wave theory in low dimensions
\cite{T2494,H4769,T5000,I1082,Y157603,K104427,H1453} that
Takahashi \cite{T168} gave a spin-wave description of the one-dimensional
ferromagnetic thermodynamics introducing an additional constraint on the
number of spin waves.
This modified spin-wave scheme was further applied to spin-gapped
antiferromagnets \cite{H549,R2589,Y769} and qualitatively improved for
one-dimensional ferrimagnets. \cite{Y14008,Y11033}
The antiferromagnetic modified spin-wave theory is less quantitative
than the ferrimagnetic version, \cite{Y064426,N214418} but it
enlighteningly interpreted novel observations such as the temperature
dependence of the Haldane gap \cite{R2589,Y769} and the field dependence
of the nuclear spin-lattice relaxation rate. \cite{Y822}
Such spin-wave understanding is well supported by other analytic
descriptions. \cite{J9265,S9188,T13515}
As for finite-temperature calculation of spin-gapped antiferromagnets,
the Schwinger-boson mean-field theory is of no use, while the modified
spin-wave theory maintains its validity to a certain extent.
\cite{Y064426}
Thus, we apply the modified spin-wave scheme to the spin-$\frac{1}{2}$
ferromagnetic-antiferromagnetic bond-alternating chain with particular
emphasis on a comparison between fermionic and bosonic descriptions of
spin-gapped antiferromagnets.

   Our theoretical attempt is much motivated by existent bond-alternating
chain compounds such as
IPACuCl$_3$
($\mbox{IPA}=\mbox{isopropylammonium}=(\mbox{CH}_3)_2\mbox{CHNH}_3$)
\cite{B125} and
(4-BzpipdH)CuCl$_3$
($\mbox{4-BzpipdH}
 =\mbox{4-benzylpiperidinium}=\mbox{C}_{12}\mbox{H}_{18}\mbox{N}$).
\cite{R2603}
These materials behave as spin-$1$ Haldane-gap antiferromagnets at low
temperatures, \cite{M564,H1792,M3913,M675} while such spin-$1$ features
are broken up into paramagnetic spin $\frac{1}{2}$'s with increasing
temperature. \cite{M14279,M144428,M2694}
Besides the thermal crossover from quantum spin $1$'s to classical spin
$\frac{1}{2}$'s, their enriched ground-state properties
\cite{H2207,H8268,K3486,Y9555,O2587} and novel edge states \cite{F220} are
of great interest to both theoreticians and experimentalists.

\section{Formalism}

   We consider the ferromagnetic-antiferromagnetic bond-dimeric
spin-$\frac{1}{2}$ Heisenberg chain, whose Hamiltonian is given by
\begin{eqnarray}
   &&
   {\cal H}
   =\sum_{n=1}^N
    \Bigl[
    \bigl(
     J_{\rm AF}\mbox{\boldmath$S$}_{2n-1}\cdot\mbox{\boldmath$S$}_{2n  }
    -J_{\rm  F}\mbox{\boldmath$S$}_{2n  }\cdot\mbox{\boldmath$S$}_{2n+1}
    \bigr)
   \nonumber\\
   &&\qquad
   -g\mu_{\rm B}H
    \bigl(S_{2n-1}^z+S_{2n}^z\bigr)
    \Bigr].
   \label{E:H}
\end{eqnarray}
The ground-state properties \cite{T428,H1879,H439,S251} and low-lying
excitations \cite{H2514} of this model were well investigated by
numerical tools and variational schemes.
In particular, the string order parameter originally defined for spin-$1$
Heisenberg chains \cite{N4709} was generalized to this system
\cite{T428,H1879,H439} and the breakdown of a hidden $Z_2\times Z_2$
symmetry was extensively argued. \cite{K3486,Y9555}
As the ferromagnetic coupling tends to infinity, the string order remains
finite and the Haldane gap converges to that originating in decoupled
singlet dimers.

   On the other hand, the thermodynamic properties have much less been
calculated so far \cite{F220,H1416} and there is no guiding theory for
extensive experimental findings.
Employing two different languages, we calculate various thermal quantities
and give rigorous information on their low-temperature behavior.

\subsection{Fermionic Approach}

   In accordance with the bond dimerization, we introduce two kinds of
spinless fermions through the Jordan-Wigner transformation
\begin{eqnarray}
   &&
   S_{2n-1}^+
    =a_n^\dagger{\rm exp}
     \Biggl[{\rm i}\pi
      \Biggl(
       \sum_{m=1}^{n-1}a_m^\dagger a_m
      +\sum_{m=1}^{n-1}b_m^\dagger b_m
      \Biggr),
   \nonumber\\
   &&
   S_{2n  }^+
    =b_n^\dagger{\rm exp}
     \Biggl[{\rm i}\pi
      \Biggl(
       \sum_{m=1}^{n  }a_m^\dagger a_m
      +\sum_{m=1}^{n-1}b_m^\dagger b_m
      \Biggr)
     \Biggr],
   \nonumber\\
   &&
   S_{2n-1}^z=a_n^\dagger a_n-\frac{1}{2},\ \ 
   S_{2n  }^z=b_n^\dagger b_n-\frac{1}{2}.
   \label{E:JW}
\end{eqnarray}
Decomposing the fermionic Hamiltonian at the Hartree-Fock level, we obtain
a mean-field Hamiltonian as
\begin{eqnarray}
   &&
   {\cal H}_{\rm HF}
    =E_0+\bigl(J_{\rm AF}-J_{\rm F}\bigr)
   \nonumber\\
   &&\qquad\times
     \sum_{n=1}^N
     \biggl[
      \Bigl(d_b-\frac{1}{2}\Bigr)a_n^\dagger a_n
     +\Bigl(d_a-\frac{1}{2}\Bigr)b_n^\dagger b_n
     \biggr]
   \nonumber\\
   &&\qquad
    +J_{\rm AF}\sum_{n=1}^N
     \biggl[
      \Bigl(\frac{1}{2}-p_{\rm AF}  \Bigr)a_n^\dagger b_n
     +{\rm H.c.}
     \biggr]
   \nonumber\\
   &&\qquad
    -J_{\rm F}\sum_{n=1}^N
     \biggl[
      \Bigl(\frac{1}{2}-p_{\rm F}  \Bigr)b_n^\dagger a_{n+1}
     +{\rm H.c.}
     \biggr]
   \nonumber\\
   &&\qquad
    -g\mu_{\rm B}H\sum_{n=1}^N
     \bigl(a_n^\dagger a_n+b_n^\dagger b_n\bigr),
   \label{E:HSFHF}
\end{eqnarray}
where
$d_a=\langle a_n^\dagger a_n\rangle_{\rm HF}$,
$d_b=\langle b_n^\dagger b_n\rangle_{\rm HF}$,
$p_{\rm AF}=\langle b_n^\dagger a_n\rangle_{\rm HF}$,
$p_{\rm F}=\langle a_{n+1}^\dagger b_n\rangle_{\rm HF}$, and
\begin{eqnarray}
   &&
   E_0
    =\biggl[
      J_{\rm AF}
      \Bigl(|p_{\rm AF}|^2-d_ad_b+\frac{1}{4}\Bigr)
   \nonumber\\
   &&\qquad
     -J_{\rm  F}
      \Bigl(|p_{\rm  F}|^2-d_ad_b+\frac{1}{4}\Bigr)
     +g\mu_{\rm B}H
     \biggr]N,
\end{eqnarray}
with $\langle\cdots\rangle_{\rm HF}$ denoting the thermal average over the
Hartree-Fock eigenstates.
Defining the Fourier transformation as
\begin{equation}
   \left.
   \begin{array}{ccc}
   a_n&=&{\displaystyle\frac{1}{\sqrt{N}}}
      {\displaystyle\sum_k}{\rm e}^{{\rm i}k(n-1/4)}a_k,
   \\
   b_n&=&{\displaystyle\frac{1}{\sqrt{N}}}
      {\displaystyle\sum_k}{\rm e}^{{\rm i}k(n+1/4)}b_k,
   \\
   \end{array}
   \right.
\end{equation}
and then a unitary transformation as
\begin{equation}
   \left(
    \begin{array}{c}
       a_k\\ b_k\\
    \end{array}
   \right)
  =\left(
    \begin{array}{cc}
     u_k & v_k{\rm e}^{ {\rm i}\theta_k} \\
     v_k{\rm e}^{-{\rm i}\theta_k} & -u_k \\
    \end{array}
   \right)
   \left(
    \begin{array}{c}
     \alpha_k\\ \beta_k\\
    \end{array}
   \right),
\end{equation}
where
\begin{eqnarray}
   &&
   u_k=\sqrt{\frac{1}{2}
       \Bigl(1-\frac{\eta}{\sqrt{\eta^2+|\gamma_k|^2}}\Bigr)},
   \nonumber\\
   &&
   v_k=\sqrt{\frac{1}{2}
       \Bigl(1+\frac{\eta}{\sqrt{\eta^2+|\gamma_k|^2}}\Bigr)},
   \nonumber\\
   &&
   \gamma_k\equiv|\gamma_k|{\rm e}^{{\rm i}\theta_k}
   \nonumber\\
   &&\quad\ 
    =J_{\rm AF}\Bigl(\frac{1}{2}-p_{\rm AF}\Bigr){\rm e}^{{\rm i}k/2}
    -J_{\rm F}\Bigl(\frac{1}{2}-p_{\rm F}^*\Bigr){\rm e}^{-{\rm i}k/2},
   \nonumber\\
   &&
   \xi=\frac{1}{2}
       \bigl(J_{\rm AF}-J_{\rm F}\bigr)\bigl(d_a+d_b-1\bigr),
   \nonumber\\
   &&
   \eta=\frac{1}{2}
        \bigl(J_{\rm AF}-J_{\rm F}\bigr)\bigl(d_a-d_b\bigr),
\end{eqnarray}
and twice the lattice constant is set equal to unity,
we can diagonalize the Hamiltonian as
\begin{equation}
   {\cal H}_{\rm HF}=E_0+\sum_k
    \bigl(
     \varepsilon_k^+\alpha_k^\dagger\alpha_k
    +\varepsilon_k^-\beta_k^\dagger\beta_k
    \bigr),
   \label{E:HSFHFdiag}
\end{equation}
where the dispersion relations are given by
\begin{equation}
   \varepsilon_k^\pm=\xi\pm\sqrt{\eta^2+|\gamma_k|^2}-g\mu_{\rm B}H.
   \label{E:dspSF}
\end{equation}

   In terms of the fermion distribution functions
$\bar{n}_k^\pm=[{\rm e}^{\varepsilon_k^\pm/k_{\rm B}T}+1]^{-1}$,
the internal energy, the total magnetization, and the magnetic
susceptibility are expressed as
\begin{eqnarray}
   &&
   E=E_0
    +\sum_k\sum_{\sigma=\pm}\varepsilon_k^\sigma\bar{n}_k^\sigma,
   \label{E:ESF}
   \\
   &&
   M=\sum_k\sum_{\sigma=\pm}\bar{n}_k^\sigma-N,
   \label{E:MSF}
   \\
   &&
   \chi=\frac{(g\mu_{\rm B})^2}{k_{\rm B}T}\sum_k\sum_{\sigma=\pm}
        \bar{n}_k^\sigma\bigl(1-\bar{n}_k^\sigma),
   \label{E:chiSF}
\end{eqnarray}
respectively.
Another quantity of wide interest is the nuclear spin-lattice relaxation
rate $1/T_1$.
Considering the electronic-nuclear energy-conservation requirement, the
Raman process usually plays a leading role in the relaxation, which is
formulated as
\begin{eqnarray}
   &&
   \frac{1}{T_1}
    =\frac{4\pi(g\mu_{\rm B}\hbar\gamma_{\rm N})^2}
          {\hbar\sum_m{\rm e}^{-E_m/k_{\rm B}T}}
     \sum_{m,m'}{\rm e}^{-E_m/k_{\rm B}T}
   \nonumber\\
   &&\qquad\times
     \Biggl|
      \Bigl\langle
       m'\Bigl|\sum_n\bigl(A_nS_{2n-1}^z+B_nS_{2n}^z\bigr)\Bigr|m
      \Bigr\rangle
     \Biggr|^2
   \nonumber\\
   &&\qquad\times
     \delta(E_{m'}-E_m-\hbar\omega_{\rm N}),
   \label{E:T1def}
\end{eqnarray}
where $A_n$ and $B_n$ are the dipolar coupling constants between
the nuclear and electronic spins, $\omega_{\rm N}\equiv\gamma_{\rm N}H$ is
the Larmor frequency of the nuclei with $\gamma_{\rm N}$ being the
gyromagnetic ratio, and the summation $\sum_m$ is taken over all the
electronic eigenstates $|m\rangle$ with energy $E_m$.
Assuming the Fourier components of the coupling constants to have little
momentum dependence as
$\sum_n e^{{\rm i}kn}A_n\equiv A_k\simeq A$ and
$\sum_n e^{{\rm i}kn}B_n\equiv B_k\simeq B$,
we obtain the fermionic expression of the Raman relaxation rate as
\begin{eqnarray}
   &&
   \frac{1}{T_1}
    =\frac{\pi(g\mu_{\rm B}\hbar\gamma_{\rm N})^2}
          {\hbar N^2}
     \sum_{k,k'}
     \bigl[
      A^2+B^2+2AB{\rm cos}(\theta_{k'}-\theta_k)
     \bigr]
   \nonumber\\
   &&\qquad\times
     \sum_{\sigma=\pm}
     \bar{n}_k^\sigma\bigl(1-\bar{n}_{k'}^\sigma)
     \delta(\varepsilon_{k'}^\sigma
           -\varepsilon_k^\sigma
           -\hbar\omega_{\rm N}).
   \label{E:T1SF}
\end{eqnarray}

\subsection{Bosonic Approach}

   Next we consider a single-component bosonic representation of each spin
variable at the cost of the rotational symmetry.
We start from the Holstein-Primakoff transformation
\begin{eqnarray}
   &&
   S_{4n-4+\tau}^+=\sqrt{2S-a_{\tau:n}^\dagger a_{\tau:n}}
                    \ a_{\tau:n},
   \nonumber\\
   &&
   S_{4n-4+\tau}^z=S-a_{\tau:n}^\dagger a_{\tau:n},
   \nonumber\\
   &&
   S_{4n-2+\tau}^+=b_{\tau:n}^\dagger
                   \sqrt{2S-b_{\tau:n}^\dagger b_{\tau:n}},
   \nonumber\\
   &&
   S_{4n-2+\tau}^z=-S+b_{\tau:n}^\dagger b_{\tau:n},
   \label{E:HP}
\end{eqnarray}
where $\tau=1,2$; that is, we assume the chain to consist of four
sublattices.
Under the large-$S$ treatment, the Hamiltonian can be expanded as
\begin{equation}
   {\cal H}=-2(J_{\rm F}+J_{\rm AF})S^2N
            +E_1+E_0+{\cal H}_1+{\cal H}_0+O(S^{-1}),
   \label{E:HSW}
\end{equation}
where $E_i$ and ${\cal H}_i$ give the $O(S^i)$ quantum corrections to the
ground-state energy and the dispersion relations, respectively.
The naivest diagonalization of the Hamiltonian (\ref{E:HSW}), whether
up to $O(S^1)$ or up to $O(S^0)$, results in diverging sublattice
magnetizations even at zero temperature.
In order to suppress the quantum as well as thermal divergence of the
number of bosons, we consider minimizing the free energy constraining the
sublattice magnetizations to be zero: \cite{T2494,H4769,T5000}
\begin{equation}
   \sum_{n=1}^N\sum_{\tau=1,2}
   \bigl(
    a_{\tau:n}^\dagger a_{\tau:n}+b_{\tau:n}^\dagger b_{\tau:n}
   \bigr)
  =4SN.
  \label{E:const}
\end{equation}
Within the conventional spin-wave theory, spins on one sublattice point
predominantly up, while those on the other predominantly down.
The condition (\ref{E:const}) restores the sublattice symmetry.
In order to enforce the constraint (\ref{E:const}), we first
introduce a Lagrange multiplier and diagonalize
\begin{equation}
   \widetilde{\cal H}
   ={\cal H}+J_{\rm AF}\nu S\sum_{n=1}^N\sum_{\tau=1,2}
   \bigl(
    a_{\tau:n}^\dagger a_{\tau:n}+b_{\tau:n}^\dagger b_{\tau:n}
   \bigr).
   \label{E:effHSW}
\end{equation}

   We define the Fourier transformation as
\begin{equation}
   \left.
   \begin{array}{ccc}
   a_{\tau:n}&=&{\displaystyle\frac{1}{\sqrt{N}}}
    {\displaystyle\sum_k}{\rm e}^{-{\rm i}k(n-5/8+\tau/4)}a_{\tau:k},
   \\
   b_{\tau:n}&=&{\displaystyle\frac{1}{\sqrt{N}}}
    {\displaystyle\sum_k}{\rm e}^{ {\rm i}k(n-1/8+\tau/4)}b_{\tau:k},
   \\
   \end{array}
   \right.
\end{equation}
and then the  Bogoliubov transformation as
\begin{equation}
   \left(
    \begin{array}{c}
     a_{1:k}         \\
     a_{2:k}         \\
     b_{1:k}^\dagger \\
     b_{2:k}^\dagger \\
    \end{array}
   \right)
  =\left(
    \begin{array}{rrrr}
     \psi_{1:k}^+      &  \psi_{1:k}^-      &
    -\psi_{4:k}^{+\,*} & -\psi_{4:k}^{-\,*} \\
     \psi_{2:k}^+      &  \psi_{2:k}^-      &
    -\psi_{3:k}^{+\,*} & -\psi_{3:k}^{-\,*} \\
    -\psi_{3:k}^+      & -\psi_{3:k}^-      &
     \psi_{2:k}^{+\,*} &  \psi_{2:k}^{-\,*} \\
    -\psi_{4:k}^+      & -\psi_{4:k}^-      &
     \psi_{1:k}^{+\,*} &  \psi_{1:k}^{-\,*} \\
    \end{array}
   \right)
   \left(
    \begin{array}{c}
     \alpha_{1:k}         \\
     \beta_{1:k}          \\
     \alpha_{2:k}^\dagger \\
     \beta_{2:k}^\dagger  \\
    \end{array}
   \right),
\end{equation}
where four times the lattice constant is set equal to unity.
We determine the coefficients $\psi_{i:k}^\pm$ so as to diagonalize
$\widetilde{\cal H}$ up to the order of $O(S)$ and perturbationally take
${\cal H}_0$ into calculation. \cite{Y211}
Then the Hamiltonian (\ref{E:HSW}) is written as
\begin{eqnarray}
   &&
   E_1=-2J_{\rm AF}(1+\gamma+\nu)SN
      +J_{\rm AF}\sum_k\sum_{\sigma=\pm}\omega_k^\sigma,
   \nonumber\\
   &&
   E_0=2J_{\rm AF}
       \bigl[
        2{\mit\Delta}({\mit\Lambda}-\gamma{\mit\Gamma})
       -(1+\gamma){\mit\Delta}^2
       -\gamma{\mit\Gamma}^2
       -{\mit\Lambda}^2
       \bigr],
   \nonumber\\
   &&
   {\cal H}_1
   =J_{\rm AF}\sum_k\sum_{\tau=1,2}
    \bigl(
     \omega_k^+\alpha_{\tau:k}^\dagger\alpha_{\tau:k}
    +\omega_k^-\beta_{\tau:k}^\dagger\beta_{\tau:k}
    \bigr),
   \nonumber\\
   &&
   {\cal H}_0
   =J_{\rm AF}\sum_k\sum_{\tau=1,2}
    \bigl(
     \delta\omega_k^+\alpha_{\tau:k}^\dagger\alpha_{\tau:k}
    +\delta\omega_k^-\beta_{\tau:k}^\dagger\beta_{\tau:k}
    \bigr)
   \nonumber\\
   &&\qquad
   +{\cal H}_{\rm irrel}+{\cal H}_{\rm quart},
\end{eqnarray}
where $\gamma=J_{\rm F}/J_{\rm AF}$,
\begin{eqnarray}
   &&
   \omega_k^\sigma
   =S\sqrt{(1+\gamma+\nu)^2-1+\gamma^2+2\sigma\chi_k},
   \nonumber\\
   &&
   \delta\omega_k^\sigma
   =\bigl[({\mit\Lambda}-\gamma{\mit\Gamma}-(1+\gamma){\mit\Delta}\bigr]
    \Bigl(
     1+\frac{\sigma\gamma}{\chi_k}
    \Bigr)
    \frac{1+\gamma+\nu}{\lambda_k^\sigma}
   \nonumber\\
   &&\qquad
   -\gamma({\mit\Delta}+{\mit\Gamma})
    \frac{\gamma+\sigma\chi_k}{\lambda_k^\sigma}
   +({\mit\Delta}-{\mit\Lambda})
    \frac{\sigma\chi_k+\gamma{\rm sin}^2\frac{k}{2}}
         {\sigma\chi_k\lambda_k^\sigma},
   \nonumber\\
   &&
   {\mit\Gamma}
   =\frac{1}{4N}\sum_k\sum_{\sigma=\pm}
    \biggl[
     \Bigl(
      1+\frac{\sigma\gamma}{\chi_k}
     \Bigr)
     \frac{1+\gamma+\nu}{\lambda_k^\sigma}-1
    \biggr]
   \nonumber\\
   &&\qquad\times
    \frac{(1+\gamma+\nu)\sigma\chi_k{\rm cos}^2\frac{k}{2}
         -\gamma(\gamma+\sigma\chi_k)\lambda_k^\sigma
          {\rm sin}^2\frac{k}{2}}
         {\chi_k^2+\gamma(\gamma+2\sigma\chi_k){\rm sin}^2\frac{k}{2}},
   \nonumber\\
   &&
   {\mit\Lambda}
   =\frac{1}{4N}\sum_k\sum_{\sigma=\pm}
    \frac{\sigma\chi_k+\gamma{\rm sin}^2\frac{k}{2}}
         {\sigma\chi_k\lambda_k^\sigma},
   \nonumber\\
   &&
   {\mit\Delta}
   =\frac{1}{4N}\sum_k\sum_{\sigma=\pm}
    \biggl[
     \Bigl(
      1+\frac{\sigma\gamma}{\chi_k}
     \Bigr)
     \frac{1+\gamma+\nu}{\lambda_k^\sigma}-1
    \biggr],
   \label{E:dspMSW}
\end{eqnarray}
with $\lambda_k^\sigma=\omega_k^\sigma/S$ and
$\chi_k=[(1+\gamma+\nu)^2-{\rm sin}^2(k/2)]^{1/2}$.
${\cal H}_{\rm irrel}$ and ${\cal H}_{\rm quart}$ in ${\cal H}_0$ contain
off-diagonal one-body terms such as $\alpha_{\tau:k}\beta_{\tau:k}$ and
residual two-body interactions, respectively, both of which are neglected
in the perturbational treatment.

   At finite temperatures we replace
$\alpha_{\tau:k}^\dagger\alpha_{\tau:k}$ and
$\beta_{\tau:k}^\dagger\beta_{\tau:k}$ by their canonical averages
$\langle\alpha_{\tau:k}^\dagger\alpha_{\tau:k}\rangle
 \equiv\bar{n}_{\tau:k}^+$ and
$\langle\beta_{\tau:k}^\dagger\beta_{\tau:k}\rangle
 \equiv\bar{n}_{\tau:k}^-$, respectively, which are expressed as
$\bar{n}_{\tau:k}^\pm\equiv\bar{n}_k^\pm
=[{\rm e}^{J_{\rm AF}(\omega_k^\sigma+\delta\omega_k^\sigma)
          /k_{\rm B}T}-1]^{-1}$.
Here the Lagrange multiplier $\nu$ is determined through
\begin{equation}
   \sum_k\sum_{\sigma=\pm}
   \Bigl(1+\frac{\sigma\gamma}{\chi_k}\Bigr)
   \frac{1+\gamma+\nu}{\lambda_k^\sigma}
   \bigl(1+2\bar{n}_k^\sigma\bigr)
  =2N(1+2S).
\end{equation}
Then the internal energy and the magnetic susceptibility are given by
\cite{T233}
\begin{eqnarray}
   &&
   E=E_{\rm g}
    +2\sum_k\sum_{\sigma=\pm}
      \widetilde{\omega}_k^\sigma\bar{n}_k^\sigma,
   \nonumber\\
   &&
   \chi=\frac{2(g\mu_{\rm B})^2}{3k_{\rm B}T}
        \sum_k\sum_{\sigma=\pm}
        \bar{n}_k^\sigma\bigl(\bar{n}_k^\sigma+1\bigr),
\end{eqnarray}
respectively, where $E_{\rm g}=-2(J_{\rm F}+J_{\rm AF})S^2N+E_1+E_0$.

\section{Calculations}

\begin{figure*}
\centering
\includegraphics[width=120mm]{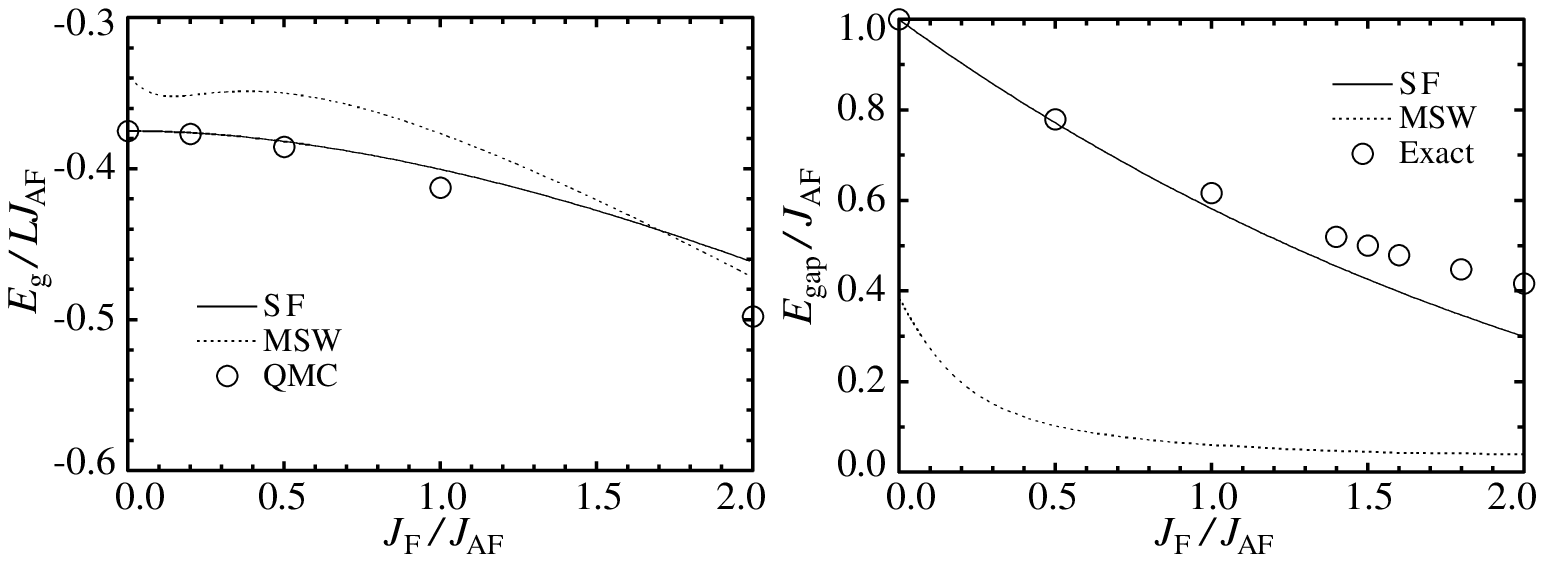}
\caption{The spinless-fermion (SF), modified-spin-wave (MSW), quantum
         Monte Carlo (QMC), and numerical-diagonalization (Exact)
         calculations of the ground-state energy (the left) and the
         excitation gap immediately above the ground state (the right) for
         the bond-alternating dimerized chain, where $L\equiv 2N$ is the
         number of spins.}
\label{F:gap}
\end{figure*}

   First we calculate the ground-state energy $E_{\rm g}$ and the lowest
excitation gap $E_{\rm gap}$ and compare them with numerical findings in
Fig. \ref{F:gap}.
The spinless fermions are much better than the modified spin waves at
describing both quantities.
As $J_{\rm F}$ goes to zero, the fermionic findings are refined and end up
with the exact values $E_{\rm g}/N=-3J_{\rm AF}/4$ and
$E_{\rm gap}=J_{\rm AF}$.
The modified spin waves considerably underestimate the spin gap.
They can not distinguish massive spin chains from massless critical
ones \cite{Y064426} to begin with, but they are still useful in
qualitatively investigating dependences of the Haldane gap on temperature
and spin quantum number. \cite{Y769}

\begin{figure*}
\centering
\includegraphics[width=160mm]{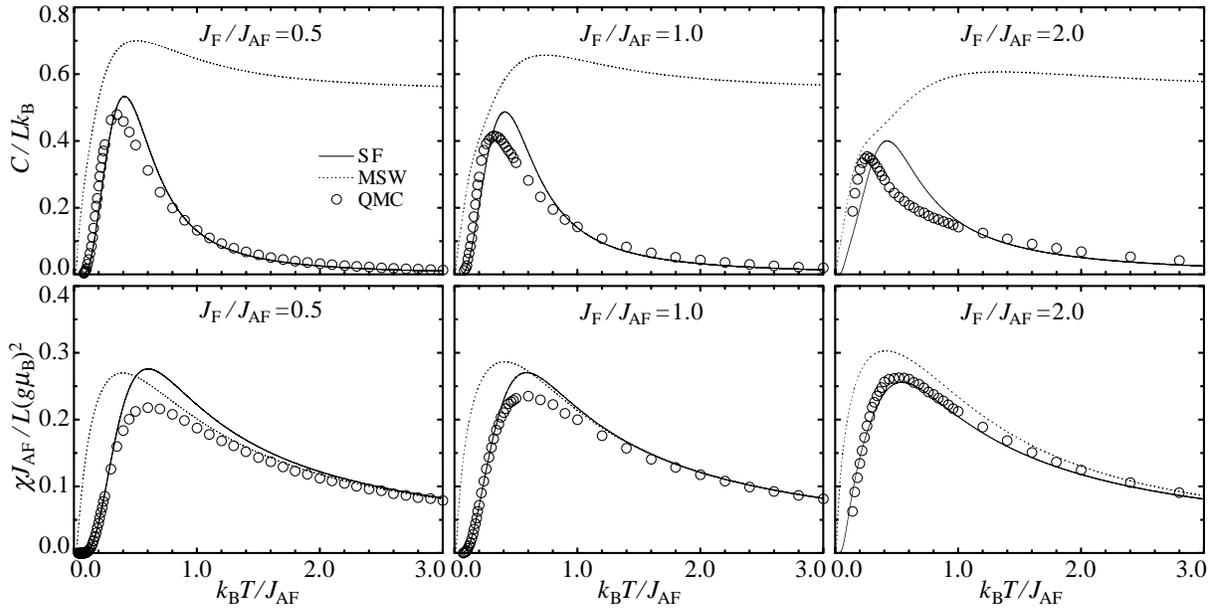}
\caption{The spinless-fermion (SF), modified-spin-wave (MSW), and quantum
         Monte Carlo (QMC) calculations of the specific heat (the upper
         three) and the magnetic susceptibility (the lower three) as
         functions of temperature for the bond-alternating dimerized
         chain, where $L\equiv 2N$ is the number of spins.}
\label{F:ThD}
\end{figure*}

   Secondly we calculate the thermodynamic properties.
Figure \ref{F:ThD} shows the temperature dependences of the zero-field
specific heat and magnetic susceptibility.
Due to the significant underestimate of the spin gap, the modified
spin-wave description is much less quantitative than the fermionic one
at low temperatures.
Furthermore, the modified spin waves completely fail to reproduce the
antiferromagnetic Schottky-type peak of the specific heat.
Because of the Lagrange multiplier $\nu$, which turns out a monotonically
increasing function of temperature, the dispersion relations
(\ref{E:dspMSW}) lead to endlessly increasing energy and thus
nonvanishing specific heat at high temperatures.
The spinless fermions succeed in reproducing the overall thermal behavior.
The present approaches have the advantage of giving the
low-temperature behavior analytically.
Equation (\ref{E:dspSF}) shows that the dispersion relation of the
low-lying excitations reads
\begin{equation}
   \varepsilon_k^\pm
   \simeq\pm(E_{\rm gap}+J_{\rm AF}vk^2)-g\mu_{\rm B}H,
   \label{E:dspSFap}
\end{equation}
provided $g\mu_{\rm B}H<E_{\rm gap}$, where
\begin{eqnarray}
   &&
   E_{\rm gap}
   =\sqrt{J_{\rm AF}^2\widetilde{p}_{\rm AF}^2
         +J_{\rm  F}^2\widetilde{p}_{\rm  F}^2
         +2J_{\rm  F}J_{\rm AF}
          \widetilde{p}_{\rm F}\widetilde{p}_{\rm AF}},
   \nonumber\\
   &&
   2E_{\rm gap}v=J_{\rm F}\widetilde{p}_{\rm F}\widetilde{p}_{\rm AF},
\end{eqnarray}
with $\widetilde{p}_{\rm F}=\mbox{Re}\,p_{\rm F}-1/2$ and
$\widetilde{p}_{\rm AF}=\mbox{Re}\,p_{\rm AF}-1/2$.
Then the low-temperature properties are calculated as
\begin{eqnarray}
   &&
   \frac{C}{Nk_{\rm B}}\simeq
    \sqrt{\frac{k_{\rm B}T}{\pi vJ_{\rm AF}}}
    {\rm e}^{-E_{\rm gap}/k_{\rm B}T}
    \biggl[
     \Bigl(\frac{E_{\rm gap}}{k_{\rm B}T}\Bigr)^2
    +      \frac{E_{\rm gap}}{k_{\rm B}T}
    +      \frac{3}{4}
    \biggr],
   \nonumber\\
   &&
   \frac{\chi J_{\rm AF}}{(g\mu_{\rm B})^2N}\simeq
    \sqrt{\frac{J_{\rm AF}}{\pi vk_{\rm B}T}}
    {\rm e}^{-E_{\rm gap}/k_{\rm B}T}.
\end{eqnarray}
These features are found in the antiferromagnetic Heisenberg two-leg
ladder as well \cite{H549,T13515} and can be regarded as common to
spin-gapped antiferromagnets.
The power-law prefactor to the activation-type temperature dependence,
which can hardly be extracted from numerical findings, is essential in
estimating the spin gap experimentally.

\begin{figure*}
\centering
\includegraphics[width=160mm]{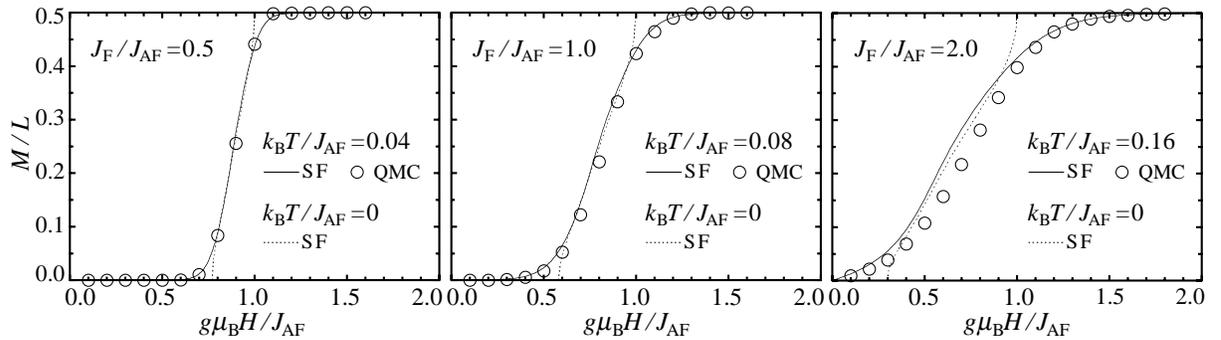}
\caption{The spinless-fermion (SF) and quantum Monte Carlo (QMC)
         calculations of magnetization curves for the bond-alternating
         dimerized chain, where $L\equiv 2N$ is the number of spins.}
\label{F:M}
\end{figure*}

   Next we consider the total magnetization as a function of an applied
field and temperature.
We compare the fermionic description of magnetization curves with
numerical findings in Fig. \ref{F:M}.
The spinless fermions again work very well.
Quantum Monte Carlo sampling becomes less and less feasible with
decreasing temperature, while we have no difficulty in calculating
Eq. (\ref{E:MSF}) even at zero temperature.
The ground-state magnetization turns out to behave as
$M\propto(H-H_{\rm c})^{1/2}$ near the critical field
$g\mu_{\rm B}H_{\rm c}=E_{\rm gap}$.\cite{S8091}
Magnetization plateaux of multi-leg spin ladders \cite{H1607} and mixed
spin chains \cite{Y5175} are also well interpreted in terms of the
spinless fermions.
On the contrary, in the modified spin-wave theory, the number of
sublattice bosons are kept constant and therefore we have no quantitative
information on the uniform magnetization as well as the staggered one.
Though the Schwinger-boson mean-field theory,
\cite{A316,Y064426,A617,S5028} which consists of a rotationally invariant
bosonic representation, still works with an applied field and/or existent
anisotropy \cite{X054419,L1131,L129} to a certain extent but rapidly loses
its validity with increasing temperature. \cite{Y064426}

\begin{figure*}
\centering
\includegraphics[width=120mm]{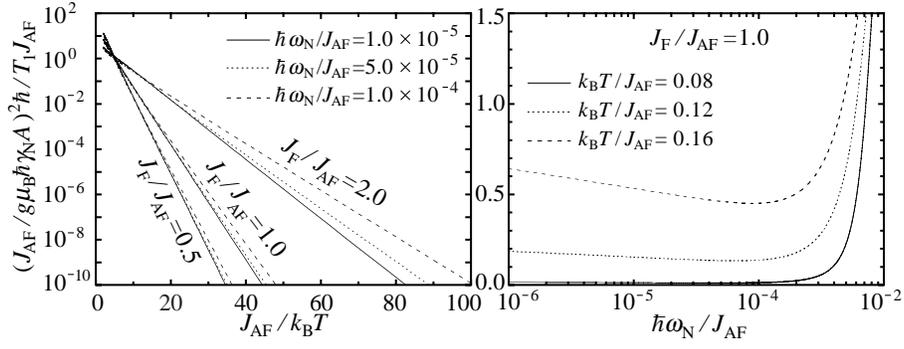}
\caption{The spinless-fermion (SF) calculations of the nuclear
         spin-lattice relaxation rate as a function of temperature
         (the left) and an applied magnetic field (the right) for the
         bond-alternating dimerized chain.}
\label{F:T1}
\end{figure*}

   Thus and thus, we are fully convinced that the spinless fermions are
superior to the modified spin waves in investigating quantum and thermal
properties of spin-gapped antiferromagnets.
Lastly in this section, we calculate the nuclear spin-lattice relaxation
rate $1/T_1$ in terms of the spinless fermions in an attempt to stimulate
further experimental interest in this system.
If we again employ the approximate dispersion (\ref{E:dspSFap}) at
moderate fields and temperatures,
$k_{\rm B}T\ll E_{\rm gap}-g\mu_{\rm B}H$, Eq. (\ref{E:T1SF}) can be
further calculated analytically as
\begin{eqnarray}
   &&
   \frac{1}{T_1}\simeq
    \frac{(g\mu_{\rm B}\hbar\gamma_{\rm N})^2}{2\pi\hbar vJ_{\rm AF}}
    (A+B)^2{\rm e}^{-E_{\rm gap}/k_{\rm B}T}
   \nonumber\\
   &&\qquad\times
    {\rm cosh}\frac{g\mu_{\rm B}H}{k_{\rm B}T}
    K_0\Bigl(\frac{\hbar\omega_{\rm N}}{2k_{\rm B}T}\Bigr),
   \label{E:T1SFap}
\end{eqnarray}
where $K_0$ is the modified Bessel function of the second kind and behaves
as $K_0(x)\simeq {\rm ln}2-\gamma-{\rm ln}x$ for $0<x\ll 1$ with $\gamma$
being Euler's constant.
Considering the significant difference between the electronic and nuclear
energy scales ($\hbar\omega_{\rm N}\alt 10^{-5}J$), there usually holds
the condition $\hbar\omega_{\rm N}\ll k_{\rm B}T$.
At low temperatures, $1/T_1$ also exhibits an increase of the activation
type but with logarithmic correction, which is much weaker than the power
correction in the case of the susceptibility.
Such a pure spin-gap-activated temperature dependence of $1/T_1$, which is
shown in Fig. \ref{F:T1}, should indeed be observed experimentally, unless
magnetic impurities mask the intrinsic properties.
Equation (\ref{E:T1SFap}) further reveals a unique field dependence of
$1/T_1$:
{\it With increasing field, $1/T_1$ first decreases logarithmically and
then increases exponentially}, which is visualized in Fig. \ref{F:T1}.
The initial logarithmic behavior comes from the Van Hove singularity
peculiar to one-dimensional energy spectra and may arise from a nonlinear
dispersion relation at the band bottom in more general.
Therefore, besides spin-gapped antiferromagnets, one-dimensional
ferromagnets and ferrimagnets may exhibit a similar field dependence.
\cite{Y822,T13515,Y2324,H054409}
Relaxation-time measurements on spin-gapped chain antiferromagnets such as
IPACuCl$_3$ and (4-BzpipdH)CuCl$_3$ are strongly encouraged.

\section{Bond-Alternating Ferrimagnetic Chain}

   Before closing our comparative study, we briefly mention a
bond-alternating but ferrimagnetic chain calculated within the same
schemes.
We take another interest in
the ferromagnetic-ferromagnetic-antiferromagnetic-antiferromagnetic
bond-tetrameric spin-$\frac{1}{2}$ Heisenberg chain, whose Hamiltonian is
given by
\begin{eqnarray}
   &&
   {\cal H}
   =\sum_{n=1}^N
    \big[
     J_{\rm AF}
     (\mbox{\boldmath$S$}_{4n-3}\cdot\mbox{\boldmath$S$}_{4n-2}
     +\mbox{\boldmath$S$}_{4n-2}\cdot\mbox{\boldmath$S$}_{4n-1})
   \nonumber \\
   &&\qquad\quad
    -J_{\rm F}
     (\mbox{\boldmath$S$}_{4n-1}\cdot\mbox{\boldmath$S$}_{4n  }
     +\mbox{\boldmath$S$}_{4n  }\cdot\mbox{\boldmath$S$}_{4n+1})
    \big]\,.
   \label{E:Htetra}
\end{eqnarray}
Cu(3-Clpy)$_2$(N$_3$)$_2$
($\mbox{3-Clpy}=\mbox{3-chloropyridine}=\mbox{C}_5\mbox{ClH}_4\mbox{N}$)
\cite{E4466} is well described by this Hamiltonian \cite{H943} and behaves
as if it is a ferrimagnet of alternating spins $\frac{3}{2}$ and
$\frac{1}{2}$. \cite{N214418}
In the conventional spin-wave scheme, the spin deviations in each
sublattice,
$\langle a_{\tau:n}^\dagger a_{\tau:n}\rangle$ and
$\langle b_{\tau:n}^\dagger b_{\tau:n}\rangle$, diverge in the
antiferromagnetic ground state but stay finite in the ferrimagnetic one.
Without quantum divergence of the sublattice magnetization, it is not
necessary to diagonalize the effective Hamiltonian (\ref{E:effHSW}).
In an attempt to keep the dispersion relations free from temperature,
we may simply diagonalise the original Hamiltonian (\ref{E:Htetra}) and
then introduce a Lagrange multiplier so as to minimize the free energy.
\cite{Y064426}
For ferrimagnets such an idea is much superior to the original
antiferromagnetic modified spin-wave scheme. \cite{T2494,H4769,T5000}

\begin{figure*}
\centering
\includegraphics[width=120mm]{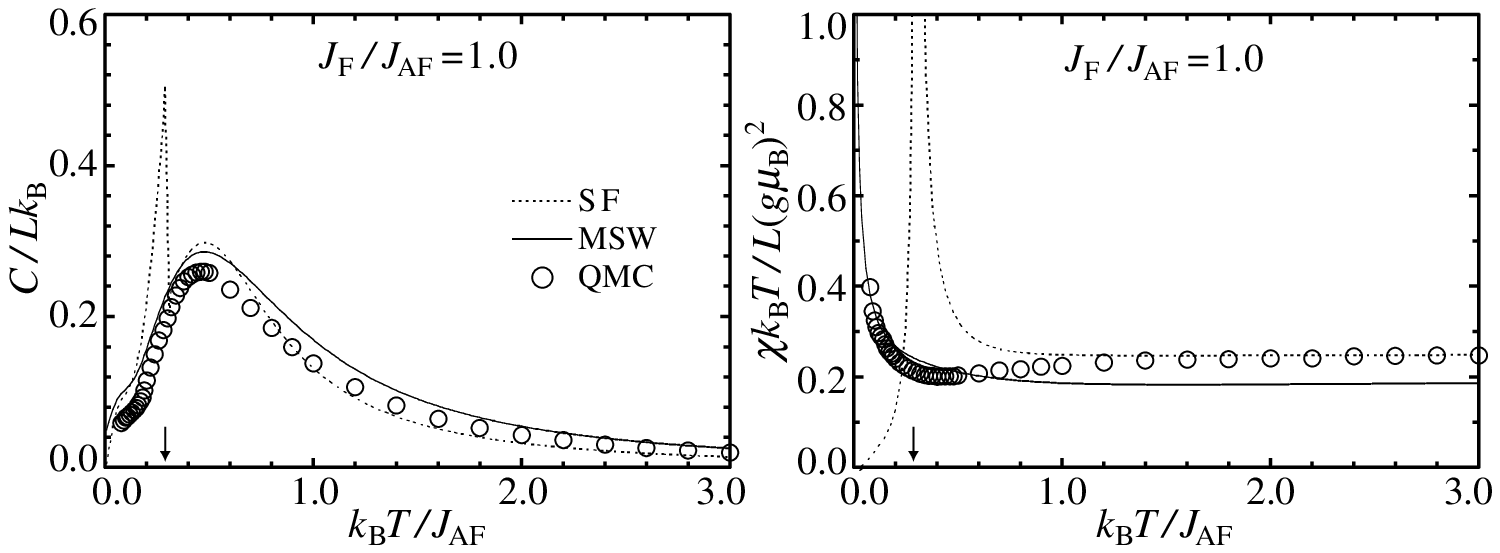}
\caption{The spinless-fermion (SF), modified-spin-wave (MSW), and quantum
         Monte Carlo (QMC) calculations of the specific heat (the left)
         and the magnetic susceptibility (the right) as functions of
         temperature for the bond-alternating tetramerized chain, where
         $L\equiv 4N$ is the number of spins.
         The Hartree-Fock fermions encounter a
         paramagnetic-to-N\'eel-ordered phase transition with decreasing
         temperature and the transition temperature is indicated by
         arrows.}
\label{F:tetra}
\end{figure*}

   Figure \ref{F:tetra} shows the thus-modified spin-wave calculations as
well as the Hartree-Fock calculations in terms of the spinless fermions
in comparison with numerical findings.
The ferrimagnetic modified spin waves work very well, contrasting with the
antiferromagnetic ones.
They reproduce the Schottky-type peak of the specific heat and the
ferrimagnetic minimum of the susceptibility-temperature product at
intermediate temperatures.
Although the modified spin-wave description of the antiferromagnetic
increase of $\chi T$ is somewhat moderate, it converges into the
paramagnetic value $S(S+1)/3$ at high temperatures.
The description is more and more refined with decreasing temperature
and is expected to be accurate at sufficiently low temperatures.
\cite{Y11033}
The $T^{1/2}$-initial behavior of $C$ and the $T^{-2}$-diverging behavior
of $\chi$ are both correctly reproduced. \cite{Y14008}
Besides static properties, $T_1$ measurements \cite{F433}
on a ferrimagnetic chain compound
NiCu(C$_7$H$_6$N$_2$O$_6$)(H$_2$O)$_3$$\cdot$2H$_2$O
was elaborately interpreted in terms of the modified spin waves.
\cite{H9023}

   On the other hand, the spinless fermions misread the low-temperature
properties of ferrimagnetic chains.
A fatally weak point of their description is the onset of a N\'eel-ordered
state.
With increasing $J_{\rm F}$, the transition temperature $T_{\rm c}$ goes
up and the applicability of the Hartree-Fock fermions is reduced.
Indeed the fermionic description is not so bad away upward from
$T_{\rm c}$, but it is much less complementary to numerical tools in
ferrimagnetic systems.

\section{Summary}

   We have comparatively discussed fermionic and bosonic descriptions of
the bond-dimeric Heisenberg chain as an example of spin-gapped
antiferromagnets.
The fermionic language is based on the Jordan-Wigner spinless fermions
within the Hartree-Fock approximation, while the bosonic formulation
consists of constraining the Holstein-Primakoff bosons to restore the
sublattice symmetry.
The spinless fermions well describe both ground-state and
finite-temperature properties.
The zero-field specific heat and magnetic susceptibility behave as
$C\propto(k_{\rm B}T)^{-3/2}{\rm e}^{-E_{\rm gap}/k_{\rm B}T}$ and
$\chi\propto(k_{\rm B}T)^{-1/2}{\rm e}^{-E_{\rm gap}/k_{\rm B}T}$,
respectively, at sufficiently low temperatures, while the relaxation rate
as
$1/T_1\propto
 {\rm e}^{-E_{\rm gap}/k_{\rm B}T}{\rm cosh}(g\mu_{\rm B}H/k_{\rm B}T)
 [0.80908-{\rm ln}(\hbar\omega_{\rm N}/k_{\rm B}T)]$
at moderately low temperatures and fields.
On the other hand, the modified spin waves give much poorer findings.
In particular, they significantly underestimate the spin gap and fail to
reproduce the Schottky-type peak of the specific heat.
The same schemes have further been applied to the bond-tetrameric
ferrimagnetic chain, where the modified spin waves work very well and are
superior to the spinless fermions both qualitatively and quantitatively.
The fermionic language is useful in describing disordered ground states
and their excitations, whereas the bosonic one in depicting ordered ground
states and their fluctuations.

   The modified spin-wave theory is fully applicable to higher-spin
systems.
The Jordan-Wigner transformation can also be generalized to higher-spin
systems, \cite{B1082} where spin-$1$ chains, for instance, are mapped onto
an extended $t$-$J$ model of strongly correlated electrons.
However, the {\it double-graded} Hubbard operators such as
$\widetilde{c}_{n,\uparrow}
 \equiv(1-c_{n,\downarrow}^\dagger c_{n,\downarrow})c_{n,\uparrow}$
demand that we should treat the fermion and boson degrees of freedom in
the same footing.\cite{A617,S2768,K5142,Y32}
The present naive fermionic representation is highly successful for
spin-$\frac{1}{2}$ gapped antiferromagnets, including various
bond-alternating and/or coupled chains.
It is complementary to numerical tools especially at low temperatures and
allows us to readily infer both static and dynamic properties of
spin-gapped antiferromagnets.

\acknowledgments

This work was supported by the Ministry of Education, Culture, Sports,
Science and Technology of Japan, and the Iketani Science and Technology
Foundation.

\begin{references}
\bibitem{H464}
   F. D. M. Haldane,
      Phys. Lett. A {\bf 93}, 464 (1983).

\bibitem{H1153}
   F. D. M. Haldane,
      Phys. Rev. Lett. {\bf 50}, 1153 (1983).

\bibitem{B371}
   W. J. L. Buyers, R. M. Morra, R. L. Armstrong, M. J. Hogan, P. Gerlach,
   and K. Hirakawa,
      Phys. Rev. Lett. {\bf 56}, 371 (1986).

\bibitem{R945}
   J. P. Renard, M. Verdaguer, L. P. Regnault, W. A. C. Erkelens,
   J. Rossat-Mignod, and W. G. Stirling,
      Europhys. Lett. {\bf 3}, 945 (1987).

\bibitem{K86}
   K. Katsumata, H. Hori, T. Takeuchi, M. Date, A. Yamagishi, and
   J. P. Renard,
      Phys. Rev. Lett. {\bf 63}, 86 (1989).

\bibitem{A799}
   I. Affleck, T. Kennedy, E. H. Lieb, and H. Tasaki,
      Phys. Rev. Lett. {\bf 59}, 799 (1987).

\bibitem{A477}
   I. Affleck, T. Kennedy, E. H. Lieb, and H. Tasaki,
      Commun. Math. Phys. {\bf 115}, 477 (1988).

\bibitem{Y3348}
   S. Yamamoto,
      Phys. Rev. Lett. {\bf 75}, 3348 (1995).

\bibitem{S493}
   U. Schollw\"ock and Th. Jolic{\oe}ur,
      Europhys. Lett. {\bf 30}, 493 (1995).

\bibitem{Y102}
   S. Yamamoto,
      Phys. Lett. A {\bf 213}, 102 (1996).

\bibitem{W14529}
   X. Wang, S. Qin, and L. Yu,
      Phys. Rev. B {\bf 60}, 14529 (1999).

\bibitem{T047203}
   S. Todo and K. Kato,
      Phys. Rev. Lett. {\bf 87}, 047203 (2001).

\bibitem{A397}
   I. Affleck,
      Nucl. Phys. B {\bf 257}, 397 (1985).

\bibitem{A409}
   I. Affleck,
      Nucl. Phys. B {\bf 265}, 409 (1986).

\bibitem{S3299}
   G. Sierra,
      J. Phys. A: Math. Gen. {\bf 29}, 3299 (1996).

\bibitem{S3443}
   G. Sierra,
      Phys. Rev. Lett. {\bf 77}, 3443 (1996).

\bibitem{F14709}
   T. Fukui and N. Kawakami,
      Phys. Rev. B {\bf 55}, R14709 (1997).

\bibitem{F2530}
   T. Fukui, M. Sigrist, and N. Kawakami,
      Phys. Rev. B {\bf 56}, 2530 (1997).

\bibitem{F8799}
   T. Fukui and N. Kawakami,
      Phys. Rev. B {\bf 56}, 8799 (1997).

\bibitem{F398}
   T. Fukui and N. Kawakami,
      Phys. Rev. B {\bf 57}, 398 (1998).

\bibitem{K622}
   A. Koga, S. Kumada, N. Kawakami, and T. Fukui,
      J. Phys. Soc. Jpn. {\bf 67}, 622 (1998).

\bibitem{T5124}
   K. Takano,
      Phys. Rev. Lett. {\bf 82}, 5124 (1999).

\bibitem{T8863}
   K. Takano,
      Phys. Rev. B {\bf 61}, 8863 (2000).

\bibitem{O1984}
   M. Oshikawa, M. Yamanaka, and I. Affleck,
      Phys. Rev. Lett. {\bf 78}, 1984 (1997).

\bibitem{T103}
   K. Totsuka,
      Phys. Lett. A {\bf 228}, 103 (1997)

\bibitem{C5126}
   D. C. Cabra, A. Honecker, and P. Pujol,
      Phys. Rev. Lett. {\bf 79}, 5126 (1997).

\bibitem{D5744}
   E. Dagotto, J. Riera, and D. Scalapino,
      Phys. Rev. B {\bf 45}, R5744 (1992).

\bibitem{G8901}
   S. Gopalan, T. M. Rice, and M. Sigrist,
      Phys. Rev. B {\bf 49}, 8901 (1994).

\bibitem{D618}
   E. Dagotto and T. M. Rice,
      Science {\bf 271}, 618 (1996).

\bibitem{K3336}
   A. K. Kolezhuk, H.-J. Mikeska, and S. Yamamoto,
      Phys. Rev, B {\bf 55}, R3336 (1997).

\bibitem{T5355}
   G.-S. Tian,
      Phys. Rev. B {\bf 56}, 5355 (1997).

\bibitem{M68}
   S. Maslov and A. Zheldev,
      Phys. Rev. B {\bf 57}, 68 (1998).

\bibitem{T15189}
   Y. Takushima, A. Koga, and N. Kawakami,
      Phys. Rev. B {\bf 61}, 15189 (2000).

\bibitem{F381}
   W.-D. Feritag and E. M\"uller-Hartmann,
      Z. Phys. B {\bf 83}, 381 (1991).

\bibitem{K281}
   A. Kl\"umper, A. Schadschneider, and J. Zittartz,
      Z. Phys. B {\bf 87}, 281 (1992).

\bibitem{K293}
   A. Kl\"umper, A. Schadschneider, and J. Zittartz,
      Europhys. Lett. {\bf 24}, 293 (1993).

\bibitem{T6443}
   K. Totsuka and M. Suzuki,
      J. Phys. A: Math. Gen. {\bf 27}, 6443 (1994).

\bibitem{T1639}
   K. Totsuka and M. Suzuki,
      J. Phys.: Condens. Matter {\bf 7}, 1639 (1995).

\bibitem{Y157}
   S. Yamamoto,
      Phys. Lett. A {\bf 225}, 157 (1997).

\bibitem{Y1795}
   S. Yamamoto,
      Int. J. Mod. Phys. B {\bf 12}, 1795 (1998).

\bibitem{L407}
   E. H. Lieb, T. Schultz, and D. J. Mattis,
      Ann. Phys. (N.Y.) {\bf 16}, 407 (1961).

\bibitem{B685}
   L. N. Bulaevski$\breve{\mbox i}$,
      Sov. Phys. JETP {\bf 16}, 685 (1963).

\bibitem{B684}
   L. N. Bulaevski$\breve{\mbox i}$,
      Sov. Phys. JETP {\bf 17}, 684 (1963).

\bibitem{K687}
   V. M. Kontorovich and V. M. Tsukernik,
      Sov. Phys. JETP {\bf 26}, 687 (1968).

\bibitem{Z181}
   A. A. Zvyagin,
      Sov. Phys. Solid State {\bf 32}, 181 (1990).

\bibitem{A6136}
   M. Azzouz,
      Phys. Rev. B {\bf 48}, 6136 (1993).

\bibitem{A6233}
   M. Azzouz, L. Chen, and S. Moukouri,
      Phys. Rev. B {\bf 50}, 6233 (1994).

\bibitem{D964}
   X. Dai and Z. Su,
      Phys. Rev. B {\bf 57}, 964 (1998).

\bibitem{H1607}
   H. Hori and S. Yamamoto,
      J. Phys. Soc. Jpn. {\bf 71}, 1607 (2002).

\bibitem{H549}
   H. Hori and S. Yamamoto,
      J. Phys. Soc. Jpn. {\bf 73}, 549 (2004).

\bibitem{A316}
   D. P. Arovas and A. Auerbach,
      Phys. Rev. B {\bf 38}, 316 (1988).

\bibitem{X054419}
   H. Xing, G. Su, S. Gao, and J. Chu,
      Phys. Rev. B {\bf 66}, 054419 (2002).

\bibitem{W1057}
   C. Wu, B. Chen, X. Dai, Y. Yu, and Z.-B. Su,
      Phys. Rev. B {\bf 60}, 1057 (1999).

\bibitem{Y064426}
   S. Yamamoto,
      Phys. Rev. B {\bf 69}, 064426 (2004).

\bibitem{C915}
   X. Y. Chen, Q. Jiang, and W. Z. Shen,
     J. Phys.: Condens. Matter {\bf 15}, 915 (2003).

\bibitem{T2494}
   M. Takahashi,
      Phys. Rev. B {\bf 40}, 2494 (1989).

\bibitem{H4769}
   J. E. Hirsch and S. Tang,
      Phys. Rev. B {\bf 40}, 4769 (1989).

\bibitem{T5000}
   S. Tang, M. E. Lazzouni, and J. E. Hirsch,
      Phys. Rev. B {\bf 40}, 5000 (1989).

\bibitem{I1082}
   V. Y. Irkhin, A. A. Katanin, and M. I. Katsnelson,
      Phys. Rev. B {\bf 60}, 1082 (1999).

\bibitem{Y157603}
   S. Yamamoto and T. Nakanishi,
      Phys. Rev. Lett. {\bf 89}, 157603 (2002).

\bibitem{K104427}
   M. Kollar, I. Spremo, and P. Kopietz,
      Phys. Rev. B {\bf 67}, 104427 (2003).

\bibitem{H1453}
   H. Hori and S. Yamamoto,
      J. Phys. Soc. Jpn. {\bf 73}, 1453 (2004).

\bibitem{T168}
   M. Takahashi,
      Phys. Rev. Lett. {\bf 58}, 168 (1987).

\bibitem{R2589}
   S. M. Rezende,
      Phys. Rev. B {\bf 42}, 2589 (1990).

\bibitem{Y769}
   S. Yamamoto and H. Hori,
      J. Phys. Soc. Jpn. {\bf 72}, 769 (2003).

\bibitem{Y14008}
   S. Yamamoto and T. Fukui,
      Phys. Rev. B {\bf 57}, R14008 (1998).

\bibitem{Y11033}
   S. Yamamoto, T. Fukui, K. Maisinger, and U. Schollw\"ock,
      J. Phys.: Condens. Matter {\bf 10}, 11033 (1998).

\bibitem{N214418}
   T. Nakanishi and S. Yamamoto,
      Phys. Rev. B {\bf 65}, 214418 (2002).

\bibitem{Y822}
   S. Yamamoto and H. Hori,
      J. Phys. Soc. Jpn. {\bf 73}, 822 (2004).

\bibitem{J9265}
   Th. Jolic{\oe}ur and O. Golinelli,
      Phys. Rev. B {\bf 50}, 9265 (1994).

\bibitem{S9188}
   J. Sagi and I. Affleck,
      Phys. Rev. B {\bf 53}, 9188 (1996).

\bibitem{T13515}
   M. Troyer, H. Tsunetsugu, and D. W\"urtz,
      Phys. Rev. B {\bf 50}, 13515 (1994).

\bibitem{B125}
   L. P. Battaglia, A. B. Corradi, G. Marcotrigiano, L. Menabue, and
   G. C. Pellacani,
      Inorg. Chem. {\bf 19}, 125 (1980).

\bibitem{R2603}
   S. A. Roberts, D. R. Bloomquist, R. D. Willett, and H. W. Dodgen,
      J. Am. Chem. Soc. {\bf 103}, 2603 (1981).

\bibitem{M564}
   H. Manaka, I. Yamada, and K. Yamaguchi,
      J. Phys. Soc. Jpn. {\bf 66}, 564 (1997).

\bibitem{H1792}
   M. Hagiwara, Y. Narumi, K. Kindo, T. Kobayashi, H. Yamakage, K. Amaya,
   and G. Schumauch,
      J. Phys. Soc. Jpn. {\bf 66}, 1792 (1997).

\bibitem{M3913}
   H. Manaka, I. Yamada, Z. Honda, H. A. Katori, and K. Katsumata,
      J. Phys. Soc. Jpn. {\bf 67}, 3913 (1998).

\bibitem{M675}
   H. Manaka, I. Yamada, N. V. Mushnikov, and T. Goto,
      J. Phys. Soc. Jpn. {\bf 69}, 675 (2000).

\bibitem{M14279}
   H. Manaka and I. Yamada,
      Phys. Rev. B {\bf 62}, 14279 (2000).

\bibitem{M144428}
   H. Manaka, I. Yamada, M. Hagiwara, and M. Tokunaga,
      Phys. Rev. B {\bf 63}, 144428 (2001).

\bibitem{M2694}
   H. Manaka, I. Yamada, Y. Miyajima, and M. Hagiwara,
      J. Phys. Soc. Jpn. {\bf 72}, 2694 (2003)f.

\bibitem{H2207}
   K. Hida,
      Phys. Rev. B {\bf 45}, 2207 (1992).

\bibitem{H8268}
   K. Hida,
      Phys. Rev. B {\bf 46}, 8268 (1992).

\bibitem{K3486}
   M. Kohmoto and H. Tasaki,
      Phys. Rev. B {\bf 46}, 3486 (1992).

\bibitem{Y9555}
   M. Yamanaka, Y. Hatsugai, and M. Kohmoto,
      Phys. Rev. B {\bf 48}, 9555 (1993).

\bibitem{O2587}
   K. Okamoto, D. Nishino, and Y. Saika,
      J. Phys. Soc. Jpn. {\bf 62}, 2587 (1993).

\bibitem{F220}
   K. Funase and S. Yamamoto,
      Phys. Lett. A {\bf 334}, 220 (2005).

\bibitem{T428}
   S. Takada,
      J. Phys. Soc. Jpn. {\bf 61}, 428 (1992).

\bibitem{H1879}
   K. Hida and S. Takada,
      J. Phys. Soc. Jpn. {\bf 61}, 1879 (1992).

\bibitem{H439}
   K. Hida,
      J. Phys. Soc. Jpn. {\bf 62}, 439 (1993).

\bibitem{S251}
   T. Sakai,
      J. Phys. Soc. Jpn. {\bf 64}, 251 (1995).

\bibitem{H2514}
   K. Hida,
      J. Phys. Soc. Jpn. {\bf 63}, 2514 (1994).

\bibitem{N4709}
   M. den Nijs and K. Rommelse,
      Phys. Rev. B {\bf 40}, 4709 (1989).

\bibitem{H1416}
   K. Hida,
      J. Phys. Soc. Jpn. {\bf 67}, 1416 (1998).

\bibitem{Y211}
   S. Yamamoto, T. Fukui, and T. Sakai,
      Eur. Phys. J. B {\bf 15}, 211 (2000).

\bibitem{T233}
   M. Takahashi,
      Prog. Theor. Phys. Suppl. {\bf 87}, 233 (1986).

\bibitem{S8091}
   T. Sakai and M. Takahashi,
      Phys. Rev. B {\bf 57}, R8091 (1998).

\bibitem{Y5175}
   S. Yamamoto and T. Sakai,
      J. Phys.: Condens. Matter {\bf 11}, 5175 (1999).

\bibitem{A617}
   A. Auerbach and D. P. Arovas,
      Phys. Rev. Lett. {\bf 61}, 617 (1988).

\bibitem{S5028}
   S. Sarker, C. Jayaprakash, H. R. Krishnamurthy, and M. Ma,
      Phys. Rev. B {\bf 40}, 5028 (1989).

\bibitem{L1131}
   C. J. De Leone and G. T. Zimanyi,
      Phys. Rev. B {\bf 49}, 1131 (1994).

\bibitem{L129}
   M.-R. Li, Y.-J. Wang, and C.-D. Gong,
      Z. Phys. B {\bf 102}, 129 (1997).

\bibitem{Y2324}
   S. Yamamoto,
      J. Phys. Soc. Jpn. {\bf 69}, 2324 (2000).

\bibitem{H054409}
   H. Hori and S. Yamamoto,
      Phys. Rev. B {\bf 68}, 054409 (2003).

\bibitem{E4466}
   A. Escuer, R. Vicente, M. S. E. Fallah, M. A. S. Goher and
   F. A. Mautner:
      Inorg. Chem. {\bf 37}, 4466 (1998).

\bibitem{H943}
   M. Hagiwara, Y. Narumi, K. Minami, K. Kindo, H. Kitazawa, H. Suzuki,
   N. Tsujii, and H. Abe,
      J. Phys. Soc. Jpn. {\bf 72}, 943 (2003).

\bibitem{F433}
   N. Fujiwara and M. Hagiwara,
      Solid State Commun. {\bf 113}, 433 (2000).

\bibitem{H9023}
   H. Hori and S. Yamamoto,
      J. Phys.: Condens. Matter {\bf 16}, 9023 (2004).

\bibitem{B1082}
   C. D. Batista and G. Ortiz,
      Phys. Rev. Lett. {\bf 86}, 1082 (2001).


\bibitem{S2768}
   Y. Suzumura, Y. Hasegawa, and H. Fukuyama,
      J. Phys. Soc. Jpn. {\bf 57}, 2768 (1987).

\bibitem{K5142}
   G. Kotliar and J. Liu,
      Phys. Rev. B {\bf 38}, 5142 (1988).

\bibitem{Y32}
   D. Yoshioka,
      J. Phys. Soc. Jpn. {\bf 58}, 32 (1989).

\end{references}
\end{document}